\documentclass[12pt]{article}
\usepackage{graphicx}

\def\pbnr{}
\def\speaker{Yan-Rui Liu}
\def\title{Heavy quark spin selection rules in S-wave meson-antimeson states}
\def\affiliation{School of Physics,
Shandong University, Jinan, Shandong 250100, China}
\def\support{yrliu@sdu.edu.cn}

\newcommand\pubnumber{\pbnr}
\newcommand\pubdate{\today}
\def\Title#1{\begin{center} {\Large #1 } \end{center}}
\def\Author#1{\begin{center}{ \sc #1} \end{center}}

\newcommand{\OnBehalf}[1]{\sbox0{#1}\ifdim\wd0=0pt
        {}
	\else
	{\\on behalf of #1}
	\fi}
\newcommand{\SupportedBy}[1]{\sbox0{#1}\ifdim\wd0=0pt
        {}
	\else
	{\footnote{#1}}
	\fi}
\def\Address#1{\begin{center}{ \it #1} \end{center}}

\newcommand\pubblock{\includegraphics[width=5cm]{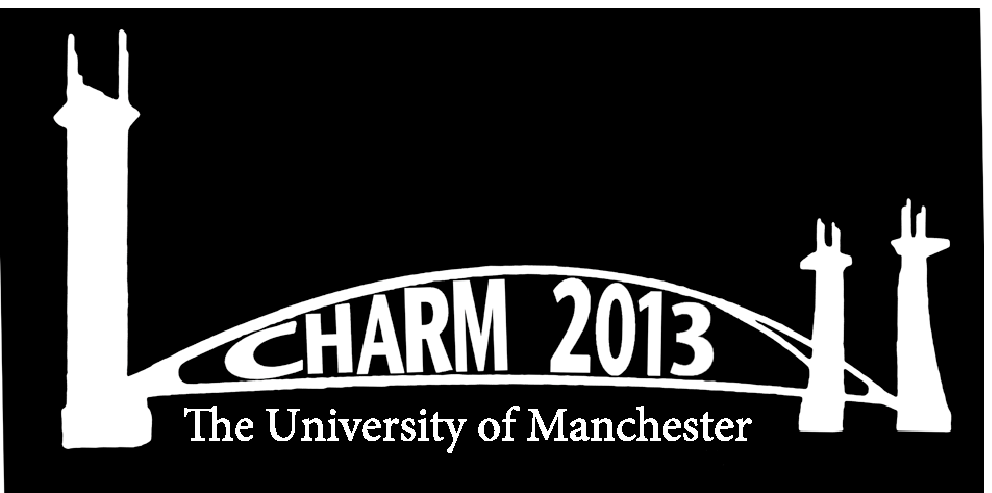}\hfill{\begin{tabular}{l} \pubnumber\\
         \pubdate  \end{tabular}}}
\newenvironment{Abstract}{\begin{quotation}  }{\end{quotation}}
\newenvironment{Presented}{\begin{quotation} \begin{center} 
             PRESENTED AT\end{center}\bigskip 
      \begin{center}\begin{large}}{\end{large}\end{center} \end{quotation}}
\def\Acknowledgements{\bigskip  \bigskip \begin{center} \begin{large}
             \bf ACKNOWLEDGEMENTS \end{large}\end{center}}
\def\venue{The 6$^{th}$ International Workshop on Charm Physics\\
(CHARM 2013)\\
Manchester, UK,  31 August -- 4 September, 2013}

\def\beq{\begin{equation}}
\def\eeq#1{\label{#1}\end{equation}}
\def\eeqn{\end{equation}}

\def\beqa{\begin{eqnarray}}
\def\eeqa#1{\label{#1}\end{eqnarray}}
\def\eeqan{\end{eqnarray}}

\let\bar=\overbar

\def\Dslash{\not{\hbox{\kern-4pt $D$}}}
\def\dslash{\not{\hbox{\kern-2pt $\del$}}}

\def\msb{{\bar{\ssstyle M \kern -1pt S}}}

\begin{document}
\begin{titlepage}
\pubblock

\vfill
\Title{\title}
\vfill
\Author{\speaker\SupportedBy{\support}}
\Address{\affiliation}
\vfill
\begin{Abstract}
By studying the spin structure of an S-wave meson-antimeson system in the heavy quark limit, we find two selection rules for the $c\bar{c}$ spin $J_{c\bar{c}}$ being only 1: (a) $J=min(J_\ell)-1$ or $J=max(J_\ell)+1$; (b) $J^C=1^+,2^-,3^+,\cdots$ if the two mesons are different but belong to the same doublet. Here $J_\ell$ is the total angular momentum of the light degree of freedom. The rules may constrain decay channels of a meson-antimeson molecule or resonance. For the recently observed $Z_c(3900)$ state, it in principle can decay into $J_{c\bar{c}}=0$ charmonium channels if it is a $D\bar{D}^*$ molecule or resonance. The non-observation of $h_c\pi$ decay might imply that this state is a tetraquark.

\end{Abstract}
\vfill
\begin{Presented}
\venue
\end{Presented}
\vfill
\end{titlepage}
\def\thefootnote{\fnsymbol{footnote}}
\setcounter{footnote}{0}
%

\section{Introduction}

\begin{figure}[htb]
\centering
\includegraphics[width=5.5in]{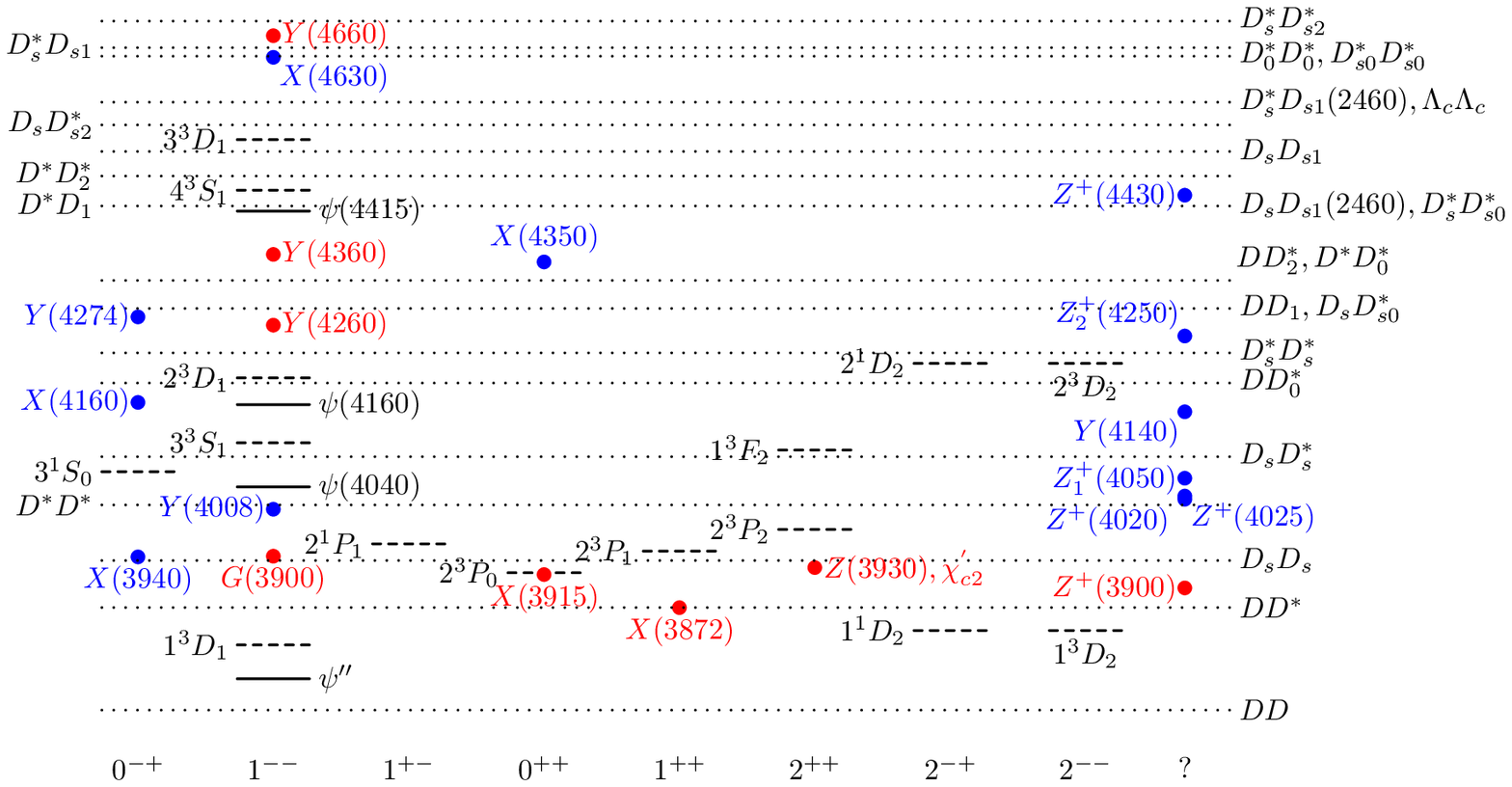}
\caption{Charmonium-like spectra and various thresholds. Dotted lines are meson-antimeson thresholds. Dashed lines are $c\bar{c}$ mesons predicted in quark model \cite{Isgur}. Solid lines are observed charmonia and solid dots are newly observed XYZ mesons. Red color means that the structure was observed by at least two experiments.}
\label{xyz}
\end{figure}

In recent years, many unexpected exotic states, which are called X, Y, or Z, are observed and parts of them have been confirmed. All the hidden charm XYZ mesons are above the $D\bar{D}$ threshold (see Fig.~\ref{xyz}), which makes their properties difficult to understand in the conventional quark model. In QCD, exotic configurations such as hadron-antihadron molecule, tetraquark, and hybrid, are allowed. It is obvious that these XYZ states are all near-threshold resonances, which can be understood in the popular molecule picture. The observation of these XYZ states provides us a good opportunity to search for meson-antimeson molecule candidates.

After the observation of three hidden-charm charged mesons $Z(4430)$, $Z_1(4050)$, and $Z_2(4250)$ by Belle \cite{charged-belle}, BESIII recently announced the charged meson $Z_c(3900)$ in Ref. \cite{Z3900-bes} and then Belle \cite{Z3900-belle} and an analysis from CLEO-c data \cite{Z3900-cleo} confirmed its existence. In the following study, BESIII reports two more charged structures $Z_c(4025)$ \cite{Z4025} and $Z_c(4020)$ \cite{Z4020} in different decay channels. The status of the spectra implies that more near-threshold heavy quark states would be observed in the future and molecules might be identified in them. It is necessary to understand the meson-antimeson picture deeper.

We here discuss a feature for meson-antimeson system: heavy quark spin selection rule, which was first noticed in Ref. \cite{Voloshin} by Voloshin. He found that the $c\bar{c}$ spin $J_{c\bar{c}}$, in the heavy quark limit, is only 1 if $X(3872)$ is a $D^0\bar{D}^{*0}$ molecule with $J^{PC}=1^{++}$. Since $X(3872)$ is also probably a charmonium with $J_{c\bar{c}}=1$, the mixing between a charmonium and a molecule is not suppressed by heavy quark mass. However, the decays into $J_{c\bar{c}}=0$ charmonia, e.g. $X(3872)\to\pi\pi\eta_c$, are suppressed. For the purpose of future investigation on molecules, we extend the study to a general meson-antimeson system. Voloshin used interpolating current and Fierz transformation to get the selection rule. We explore the problem in a different framework.

\section{Formalism}

We want to extract information for $J_{c\bar{c}}\neq0$ in the heavy quark limit. For our purpose, we construct spin wave function for a meson-antimeson system at quark level and use the re-coupling scheme to get the spin wave function of the $c\bar{c}$ pair \cite{ours}.

In the heavy quark limit, the total spin of a $c\bar{q}$ meson $j$ may be divided into two parts: the heavy quark spin $j_c=\frac12$ and the rest angular momentum $j_\ell$. $j_\ell$ is the total angular momentum of the light quark spin $s_{\bar{q}}$ and the orbital angular momentum of the system. One obviously has $\vec{j}=\vec{j}_c+\vec{j}_\ell$. In the heavy quark limit, both $j_c$ and $j_\ell$ are "good" quantum numbers. It is convenient for us to treat the system as an S-wave meson of a charm quark and an effective light antiquark $\bar{\tilde{q}}$. Then the wave function for a $D$ meson is
\begin{eqnarray}
D_j&\sim [c_{j_1}\bar{\tilde q}_{j_2}-(-1)^{j-j_1-j_2}\bar{\tilde q}_{j_2}c_{j_1}],
\end{eqnarray}
where subscripts are spins and the factor $(-1)^{j-j_1-j_2}$ comes from the exchange of two fermions. For an antimeson, we use the wavefunction
\begin{eqnarray}
\bar{D}_j&\sim [\bar{c}_{j_1}{\tilde q}_{j_2}-(-1)^{J-j_1-j_2}{\tilde q}_{j_2}\bar{c}_{j_1}].
\end{eqnarray}
Here, we have assumed the C-parity transformation to be
$$ \hat{C}: D_j\leftrightarrow \bar{D}_j .$$

Now we discuss the spin wave function of the system formed by a meson $A$ ($B$) and an antimeson $\bar{B}$ ($\bar{A}$). There are two combinations $A\bar{B}+B\bar{A}$ and $A\bar{B}-B\bar{A}$, which follows the fact that $A$ and $\bar{B}$ do not have definite C-parity while their combinations may have. Considering the exchange of two states, one gets the wave function for the system with a given C-parity $C_X$
\begin{eqnarray}\label{WF-G}
X_J&\sim&[A\bar{B}+(-1)^{J-J_A-J_B}\bar{B}A]+C_X[(-1)^{J-J_A-J_B}B\bar{A}+\bar{A}B]\nonumber\\
&\to&\Big\{(c_{j_1}\bar{q}_{\bar{j}_2})_{J_{12}} (\bar{c}_{\bar{j}_3}\tilde{q}_{j_4})_{J_{34}}+C_X(-1)^{J-J_{12}-J_{34}}(c_{j_3}\bar{\tilde{q}}_{\bar{j}_4})_{J_{34}}
(\bar{c}_{\bar{j}_1}q_{j_2})_{J_{12}}\Big\},
\end{eqnarray}
where the factor $(-1)^{J-J_A-J_B}$ ($(-1)^{J-J_{12}-J_{34}}$) is from the exchange of two bosons. We have added a bar to the spin of the antiquark in order to make it identifiable just from the symbol $\bar{j}$ in the spin wave function. It is enough for us to consider only the following part
\begin{eqnarray}\label{wf-chi}
\chi_J&=&\frac{1}{\sqrt2}\Big\{(j_1\bar{j}_2)_{J_{12}} (\bar{j}_3j_4)_{J_{34}}+C_X(-1)^{J-J_{12}-J_{34}}(j_3\bar{j}_4)_{J_{34}}
(\bar{j}_1j_2)_{J_{12}}\Big\}\nonumber\\
&=&\frac{1}{\sqrt2}\sum_{J_{13},J_{24}}(j_1\bar{j_3})_{J_{13}}[(\bar{j_2}j_4)_{J_{24}}-C_X(-1)^{j_2+j_4+J_{13}+J_{24}}(\bar{j_4}j_2)_{J_{24}}]\nonumber\\
&&\times\sqrt{(2J_{12}+1)(2J_{34}+1)(2J_{13}+1)(2J_{24}+1)}\left\{\begin{array}{ccc}\frac12&j_2&J_{12}\\\frac12&j_4&J_{34}\\J_{13}&J_{24}&J\end{array}\right\}.
\end{eqnarray}
Here, $(j_3\bar{j_1})_{J_{13}}=(j_1\bar{j_3})_{J_{13}}$ is the spin wave function of the $c\bar{c}$ pair and the part in the brackets is the spin wave function of $\bar{q}\tilde{q}$ with the C-parity $c_q=C_X(-1)^{J_{13}}$. One should note the case for $A=B$, where an additional factor $\frac{1}{\sqrt2}$ is needed and $(\bar{j_2}j_4)_{J_{24}}=(\bar{j_4}j_2)_{J_{24}}$. With this formula, it is not difficult to find the $c\bar{c}$ spin and corresponding amplitudes.

There are also suggestions that the XYZ mesons are baryon-antibaryon bound states \cite{qiao}. The above formalism may also be applied to this case. The difference lies in the light degree of freedom. One may consult details given in Ref. \cite{ours}.

\section{Heavy quark spin selection rules}

When one applies the above formula to the $D\bar{D}^*$ system, we have
\begin{eqnarray}
\chi_J=\left\{\begin{array}{lll}
(j_1\bar{j}_3)_1(\bar{j}_2j_4)_1, &&(C_X=+)\\
\frac{1}{\sqrt2}(j_1\bar{j}_3)_0(\bar{j}_2j_4)_1-\frac{1}{\sqrt2}(j_1\bar{j}_3)_1(\bar{j}_2j_4)_0, &&(C_X=-)
\end{array}\right..
\end{eqnarray}
Therefore the $c\bar{c}$ spin is only 1 for the case $J^C=1^+$, which is the result obtained in Ref. \cite{Voloshin}. By applying the formula to various hidden charm meson-antimeson systems, one finds cases for $J_{c\bar{c}}\neq0$, which are presented in Table~\ref{HQSR}. These results imply two selection rules:

(a) If $J=|j_2-j_4|-1$ or $J=j_2+j_4+1$, $J_{c\bar{c}}\neq0$;

(b) If $A$ and $B$ are different but belong to the same doublet, $J_{c\bar{c}}\neq0$ when $J^C=1^+,2^-,3^+,\cdots$.

These rules are not difficult to understand by exploring the resulting $J^C$ of the system from the two cases $J_{c\bar{c}}=1$ and $J_{c\bar{c}}=0$. The latter case gives less $J^C$ combinations. The reason that the C-parity appears in the second rule is that the light degree part has a definite C-parity.

\begin{table}[htb]
\centering
\begin{tabular}{@{}cccc@{}} \hline
State &$J^{C}$&$J_{24}$&Selection rule for $S_{c\bar{c}}\neq0$\\
$D\bar{D}^*/D_0^*\bar{D}_1^\prime$&$1^{\pm}$&$0,1$&$J^C=1^+$\\
$D^*\bar{D}^*/D_1^\prime\bar{D}_1^\prime$&$0^+,1^-,2^+$& $0,1$&$J=2$\\\hline
$D^*\bar{D}_1^\prime$&$0^\pm,1^\pm,2^\pm$& $0,1$&$J=2$\\\hline
$D^*\bar{D}_1/D_1^\prime \bar{D}_1$&$0^\pm,1^\pm,2^\pm$& $1,2$&$J=0$\\
$D^*\bar{D}^*_2/D_1^\prime\bar{D}^*_2$&$1^\pm,2^\pm,3^\pm$& $1,2$&$J=3$\\\hline
$D_1\bar{D}^*_2$&$1^\pm,2^\pm,3^\pm$&$0\sim3$&$J^C=1^+,2^-,3^+$\\
$D^*_2\bar{D}^*_2$&$0^+,1^-,2^+,3^-,4^+$&$0\sim3$&$J=4$\\ \hline
\end{tabular}
\caption{S-wave meson-antimeson states, quantum numbers, and selection rules for $J_{c\bar{c}}\neq0$.}\label{HQSR}
\end{table}

\section{Constraints for strong decays}

In the heavy quark limit, the spin-flip of the $c\bar{c}$ pair in a meson-antimeson state is suppressed, i.e. the $c\bar{c}$ spin $J_{c\bar{c}}$ is conserved in the strong decay of the system. If it contains only $J_{c\bar{c}}=1$ part, one in principle cannot observe decay products involving spin-singlet charmonium. Therefore, decay channels containing charmonium may reflect the spin structure of the initial meson.

For the interesting $D\bar{D}^{*}$ state, the possibility $I^G(J^{PC})=0^+(1^{++})$ is related with $X(3872)$ while the case $I^G(J^{PC})=1^+(1^{+-})$ is related with the newly observed $Z_c(3900)$. If $Z_c(3900)$ is a $D\bar{D}^{*}$ molecule or resonance, there is no special selection for the $c\bar{c}$ spin. Its decays into $\chi_{c0,1,2}\pi\pi$, $\eta_c\rho$, and $h_c\pi$ are all expected. However, in a recent search for $Z_c(3900)$ in the process $e^+e^-\to\pi\pi h_c$ at BESIII \cite{Z4020}, no significant signal was observed. This fact implies that $Z_c(3900)$ might contain only $J_{c\bar{c}}=1$ component and might be a compact tetraquark. If it is the case, the decay into $\eta_c\rho$ should also be suppressed. More experimental information is needed to get a stronger conclusion.

For the $D^*\bar{D}^*$ system, the allowed $J^{PC}$ are $0^{++}$, $1^{+-}$, and $2^{++}$. Here we concentrate only on isovector case which is related with charged states around the threshold. The possible decay channels for S-wave $D^*\bar{D}^*$ molecule or resonance are $0^{++}\to \eta_c\pi,\,J/\psi\rho,\,\chi_{c1}\pi$, $1^{+-}\to \eta_c\rho,\,h_c\pi,\,J/\psi\pi$, and $2^{++}\to J/\psi\rho, \, \chi_{c1}\pi,\,\chi_{c2}\pi$. BESIII observed a charged $Z_c(4020)$ in the invariant mass of $h_c\pi$ \cite{Z4020}. If it is a $D^*\bar{D}^*$ molecule, its $J^{PC}=1^{+-}$ are favored and its decay into both $\eta_c\rho$ and $J/\psi\pi$ channels should be observed. If $J/\psi\pi$ channel cannot be observed, the state might contain only $J_{c\bar{c}}=0$ component and not be a molecule. In either case, the channel $\eta_c\rho$ is expected.

Interesting observations also exist for decay properties of other meson-antimeson states. For example, if a $J^{PC}=1^{--}$ $D^*\bar{D}_2^*$ molecule or resonance exists, its branching ratios for $h_c\eta$ ($h_c\pi$) and $\eta_c\omega$ ($\eta_c\rho$) channels are expected to be large since the coefficient for the $J_{c\bar{c}}=0$ spin component in the wave function is $\frac{\sqrt{10}}{4}$.

To summarize, we discussed in what condition the $c\bar{c}$ spin is only 1 for an S-wave meson-antimeson state. We obtain two selection rules in the heavy quark limit. Since the $c\bar{c}$ spin is conserved in the strong decay, the rules are helpful to understand the structure of the initial meson from their decay channels. In reality, the mass of heavy quark is finite and the selection rules are violated. The strength of violation needs to be investigated in the future.

\Acknowledgements
This project was supported by the National Natural Science Foundation of China (No. 11275115), SRF for ROCS, SEM, and Independent Innovation Foundation of Shandong University.

\end{document}